\documentstyle[12pt]{article}
\newcommand{\be}{\begin{equation}}
\newcommand{\ee}{\end{equation}}
\begin{document}
\title{\bf Massive photons in particle and laser physics}
\author{Miroslav Pardy\\
Institute of Plasma Physics ASCR\\
Prague Asterix Laser System, PALS\\
Za Slovankou 3, 182 21 Prague 8, Czech Republic\\
and\\
Department of Physical Electronics \\
Laboratory of Plasma Physics\\
Masaryk University \\
Kotl\'{a}\v{r}sk\'{a} 2, 611 37 Brno, Czech Republic\\
e-mail:pamir@physics.muni.cz}
\date{\today}
\maketitle

\vspace{50mm}

\begin{abstract}
This article applies the theory of massive electrodynamics
to the Dirac equation with the aim to find
the generalized Volkov solution with massive photon
field. The resulting equation is the Riccati equation which cannot be
solved in general. We use the approximative Volkov
function for massive photons and consider an electron in the
periodic field and in the laser pulse of the $\delta$-function form.
We derive the modified Compton formulas for the interaction of the
multiphoton object with an electron for both cases.
\end{abstract}

\vspace{5mm}


\newpage

\baselineskip 15 pt

\section{Introduction}

The introduction of the massive photon into field theory is
elementary from the mathematical point of view. However, the physical reasons
for such generalization require serious motivation.

We know from the special
theory of relativity, that the relativistic mass formula

$$m = \frac {m_{0}}{\sqrt{1 - \frac {v^{2}}{c^{2}}}},\eqno(1)$$
where $m_{0}$ is the rest mass, has physical meaning for $v = c$, only if
$m_{0} = 0$. Since the velocity of photon in vacuum is $v = c$, it follows
from the view point of the special theory of relativity that the rest mass of
photon is zero.

Nevertheless, massless photon has a momentum

$$p = \frac {E}{c} = \frac {\hbar \omega}{c},\eqno(2)$$
as it follows from the Einstein relativistic mass formula $E =
\sqrt{c^{2}p^{2} + m^{2}c^{4}}$ in which we put the zero rest
mass of photon. Only moving photon has mass as follows from the
Einstein formula $E = mc^{2}$. Mass of the moving photon is $m_{\gamma} = \hbar
\omega/c^{2}$. A non-zero photon mass would have several implications,
such as a frequency-dependent speed of light and the existence of
longitudinal electromagnetic waves.
Photon with the nonzero rest mass is evidently in contradiction with
special relativity. Arnold Sommerfeld (1954), who first considered
superluminal velocities and theoretically discovered
the \v Cerenkov effect, wrote no remark on the massive photons
in his famous Optics.

If we suppose that the momentum of massive photon is $p = \hbar\omega/c$,
then from the Einstein formula follows that the energy of massive photon is
$E = \sqrt{\hbar^{2}\omega^{2} + m^{2}c^{4}}$.

The corresponding Planck formula for the density $P(\omega)$ of the
black body radiation is as follows:

$$P(\omega) = \left(\frac {\omega^{2}}{\pi^{2}c'^{3}}\right)
\frac {E}{e^{\frac {E}{kT}} - 1};\quad
E = \sqrt{\hbar^{2}\omega^{2} + m^{2}c^{4}},\eqno(3)$$
where we used the frequency $\omega$ instead of the momentum of photon,
 because
the frequency is used in experiment and not momentum of photon. The
massless limit of the formula (3) is the original Planck law.
We show later that $E$ given by eq. (3) is in harmony with the quantum
definition of massive photon. The quantity $c'$ is the velocity of photons
inside the black body and it must be involved in the number of
electromagnetic modes inside the blackbody. We can put approximately
$c' \approx c$.
To our knowledge there is no experimental evidence that the modified
Planck law is correct. It means that massive photons
cannot be involved into the theory of the black body radiation.
To our knowledge the precise measurement of the the anomalous magnetic moment
of electron and Lamb shift agree with QED formulas with zero photon rest
mass. On the other hand,
if photons are moving in electromagnetic field,
then they have nonzero rest mass (Ritus, 1969). This mass is complex
quantity, while we will consider here only real quantity. It follows from
the polarization operator in external fields. This operator
substantially differs from the operator in the dielectric
medium (Pardy, 1994), where the fundamental role
plays the index of refraction. Polarization of vacuum  can be
determined also by the source theory methods (Dittrich, 1978).
The photon mass following from the vacuum
polarization is not generated by the Higgs mechanism, or by the Schwinger
mechanism. This mass is of the dynamical origin corresponding to the
radiative corrections.
To our knowledge, the experiments with the black body radiation in
magnetic or electric field was never performed.

The formal introduction of the rest mass of photon exist in quantum
electrodynamics, where for instance the processes with soft photons
are calculated. In these calculations the photon mass is introduced
in order to avoid the infrared divergences (Berestetzkii {\it et al.} 1989).

We shall see later that introducing the nonzero photon mass modifies
Coulomb law. Such modification is discussed in literature (Okun,
1981). It is evident that massive photons play crucial role in
gravity. However, this problem was not discussed in the prestige articles
(Okun, 2002).

On the other hand,
the possibility that photon may be massive particle has been treated
by many physicists. The discussion is also
devoted to the existence of the mass of neutrino and its oscillations,
which can form some analogue with the photons with the same importance.
The established fact is that the massive
electrodynamics is a perfectly consistent classical and
quantum field theory (Feldmann {\it et al.}, 1963; Minkowski {\it et
  al.},
 1971;
Goldhaber {\it et al.}, 1971).
In all respect the quantum version has the same status as the standard QED.
In this article we do not solve the radiative problems in sense of article by
Nieuwenhuizen (1973). Our goal is to determine the Volkov solution of the
Dirac equation with massive photons. The resulting
equation is the Riccati equation which cannot be solved in general.
So, we derive only some approximative formulas.

In particle physics and quantum field theory
(Ryder, 1985; de Wit {\it et al.}, 1986; Commins {\it et al.}, 1983)
photon is defined
as a massless particle with spin 1. Its spin is along or in opposite direction
to its
motion. The massive photon as a neutral massive particle
is usually called vector boson. The equation for vector boson was derived in
the unified theory of the electro-weak interaction.
There are other well known examples of massive
spin-1 particles. For instance neutral $\varrho$-meson, $\varphi$-meson and
$J/\psi$ particle, bosons $W^{\pm}$ and $Z^{0}$ in particle physics.

While massless photon is described by the Maxwell
Lagrangian, the massive photon is described by the Proca Lagrangian from
which the field equations follow. The massive electrodynamics can be
considered as a generalization of massless electrodynamics.
The well known area where the massive photon or boson plays
substantial role is
the theory  of superconductivity (Ryder, 1985), plasma physics
 (Anderson, 1963),
waveguides and so on. Of course the mass of photon is not the relativistic
vacuum rest mass, but effective mass which is generated by the physical
properties of medium, or by some mechanism such as the Higgs mechanism,
Schwinger mechanism  and so on. In this sense, the physics of massive photon is
meaningful, the generalized Volkov solution of the Dirac
equation with massive photon field is physically meaningful too
and it is worthwhile to investigate problems with the massive photons.

In order to be pedagogically clear, following author's article (Pardy, 2002),
we treat,
in section 2, the massive spin 1 quantum field theory and in
section 3, we derive
the Volkov solution of the Dirac equation for massless photon field.
In section 4, we find the Riccati equation which involves mass of
photon. Then we discuss the emission of massive photons by electron in the
periodic and $\delta$-form electromagnetic field. We derive generalized Compton
formulas for interaction of the multiphotonic object with electron.

\section{Massive fields with spin 1}

If spin zero particles and fields are described by the scalar source,
then a vector source denoted here as $J^{\mu}(x)$ can be considered
as a candidate for the description of the spin 1 fields and
particles. However, there exist some obstacles because source
$J^{\mu}(x)$ has four components and spin one particles have only
three spin possibilities.

It was shown
by Schwinger (1970) that the action of the spin 1 massive particles is
as follows:

$$W(J) = \frac {1}{2}\int (dx)(dx')
\left\{J_{\mu}(x)\Delta_{+}(x-x')J^{\mu}(x') + \frac {1}{m^2}
\partial_{\mu}J^{\mu}(x)\Delta_{+}(x-x')\partial_{\nu}'J^{\nu}(x')\right\}.
\eqno(4)$$

The field of spin one particles can be defined using the definition
of the test source $\delta J^{\mu}(x)$ by the relation

$$\delta W(J) = \int\;(dx) \delta J^{\mu}(x)\varphi_{\mu}(x), \eqno(5)$$
where $\varphi_{\mu}$ is the field of particles with spin 1. After
performing variation of the formula (4) and comparison with eq. (5)
we get the equation for field of spin 1 in the following form:

$$\varphi_{\mu}(x) = \int\; (dx')\Delta_{+}(x-x')J_{\mu}(x') - \frac {1}{m^2}
\partial_{\mu}\int\; (dx')\Delta_{+}(x-x')\partial_{\nu}'J^{\nu}(x').
\eqno(6)$$

The divergence of the vector field $\varphi_{\mu}(x)$ is given by the
relation

$$\partial_{\mu}\varphi^{\mu}(x) = \int\; (dx')\Delta_{+}(x-x')
\partial_{\mu}'J^{\mu}(x') -
\frac {1}{m^2}\partial^2 \int (dx')\Delta_{+}(x-x')
\partial_{\nu}'J^{\nu}(x') = $$

$$\frac {1}{m^2}\partial_{\mu}J^{\mu}(x), \eqno(7)$$
as a consequence of scalar equation

$$-\partial^2\Delta_{+} = \delta(x-x') - m^2\Delta_{+}. \eqno(8)$$.

Further, we have after applying operator $(-\partial ^2 + m^2)$ on
the equation (6) the following equations:

$$\left(-\partial ^2 + m^2\right)\varphi_{\mu}(x) =
J_{\mu}(x) - \frac {1}{m^2}\partial_{\mu}\partial_{\nu}J^{\nu}(x),
\eqno(9)$$

$$\left(-\partial ^2 + m^2\right)\varphi_{\mu}(x) +
\partial_{\mu}\partial_{\nu}\varphi^{\nu}(x) = J_{\mu}(x),
\eqno(10)$$
as a consequence of eq. (7).

It may be easy to cast the last equation into the following form

$$\partial^{\nu}G_{\mu\nu} + m^2\varphi_{\mu} = J_{\mu}, \eqno(11)$$
where

$$G_{\mu\nu}(x) = -G_{\nu\mu}(x) =
\partial_{\mu}\varphi_{\nu} - \partial_{\nu}\varphi_{\mu}. \eqno(12)$$

Identifying $G_{\mu\nu}$ with $F_{\mu\nu}$ of the electromagnetic
field we get instead of eq. (10) and eq. (11) so called
the Proca equation for the electromagnetic field with the massive photon.

$$\left(-\partial ^2 + m^2\right)A_{\mu}(x) +
\partial_{\mu}\partial_{\nu}A^{\nu}(x) = J_{\mu}(x), \eqno(13)$$

$$\partial^{\nu}F_{\mu\nu} + m^2A_{\mu} = J_{\mu}, \eqno(14)$$

$$F_{\mu\nu}(x) = -F_{\nu\mu}(x) =
\partial_{\mu}A_{\nu} - \partial_{\nu}A_{\mu}. \eqno(15)$$

In case $m^{2} \ne 0$, we can put $\partial_{\mu}A^{\mu} = 0$ in order
to get:

$$\left(-\partial ^2 + m^2\right)A_{\mu}(x) = 0, \quad
\partial_{\mu}A^{\mu} = 0. \eqno(16)$$

The solution of the system (16) is the plane wave

$$ A_{\mu} = \varepsilon_{\mu}({\bf k})e^{ikx}, \quad k^{2} = -m^{2}
\eqno(17)$$
with $k\varepsilon({\bf k}) = 0$, which is precisely the correct definition
of a massive particle with spin 1. Later, we  will see
how to generalize this procedure
to the situation of the massive electrodynamics in dielectric
and magnetic media.

The equation (11) can be derived also from the action

$$W = \int (dx)\left(J^{\mu}(x)\varphi_{\mu}(x) +  {\cal L}
(\varphi(x))\right),
\eqno(18)$$
where

$$ {\cal L} = -\frac {1}{2}\left(\frac {1}{2}
(\partial^{\mu}\varphi^{\nu} - \partial^{\nu}\varphi^{\mu})
(\partial_{\mu}\varphi_{\nu} - \partial_{\nu}\varphi_{\mu}) +
m^{2}\varphi^{\mu}\varphi_{\mu}\right),\eqno(19)$$
where we have used the arrangement

$$\int (dx)\varphi^{\mu}(-\partial^{2})\varphi_{\mu} = \int (dx)
\partial^{\nu}\varphi^{\mu}\partial_{\nu}\varphi_{\mu} \eqno(20)$$
and

$$\int (dx)\varphi^{\mu}\partial_{\mu}\partial^{\nu}\varphi_{\nu} =
- \int (dx)\varphi^{\nu}\partial^{\mu}\partial_{\mu}\varphi_{\nu} =
- \int (dx)\varphi_{\mu}\partial^{\mu}\partial^{\nu}\varphi_{\nu}.
\eqno(21)$$

Using the last equation (21) we get the Lagrange function in the following
standard form:

$$ {\cal L} = -\frac{1}{2}
\left(\partial^{\nu}\varphi^{\mu}\partial_{\nu}\varphi_{\mu} -
(\partial_{\mu}\varphi^{\mu})^{2} + m^{2}\varphi^{\mu}\varphi_{\mu}\right).
\eqno(22)$$

If we use the A- and F-symbols, we receive from eq. (19) the Proca
Lagrangian

$$ {\cal L} = -\frac {1}{2}\left(\frac {1}{2}F^{\mu\nu}F_{\mu\nu} +
m^{2}A^{\mu}A_{\nu}\right),\eqno(23)$$
or,

$$ {\cal L} =
-\frac {1}{2}\left(\partial^{\nu}A^{\mu}\partial_{\nu}A_{\mu} -
(\partial_{\mu}A^{\mu})^{2} + m^{2}A^{\mu}A_{\mu}\right).
\eqno(24)$$

By variation of the corresponding Lagrangian for the massive field
with spin 1 we get evidently the massive Maxwell equations.

It is evident that the zero mass limit does not exist for
$\partial_{\mu}J^{\mu}(x) \not = 0$. In such a way we are forced to
redefine action $W(J)$. One of the possibilities is to put

$$\partial_{\mu}J^{\mu}(x) = mK(x) \eqno(25)$$
and identify $K(x)$ in the limit $m\rightarrow 0$ with the source of
massless spin zero particles. Since the zero mass particles with zero
spin are experimentally unknown in any event, we take $K(x) = 0$ and
we write

$$W_{[m=0]}(J) = \frac {1}{2}\int\; (dx)(dx')J_{\mu}(x)D_{+}(x-x')
J^{\mu}(x'), \eqno(26)$$
where
$$\partial_{\mu}J^{\mu}(x) = 0 \eqno(27)$$
and
$$D_{+}(x-x') = \Delta_{+}(x-x'; m=0). \eqno(28)$$

In case we want to work with electrodynamics in medium it is necessary
to involve such parameters as velocity of light $c$, magnetic permeability
$\mu$ and the dielectric constant $\varepsilon$. Then the corresponding
equations for electromagnetic potentials which are compatible with the
Maxwell equations are as follows (Schwinger {\it et al.}, 1976):

$$\left(\Delta-\frac{\mu\epsilon}{c^2}
\*\frac{\partial^2}{\partial\*t^2} + \frac{m^{2}c^{2}}{\hbar^{2}}
 \right)A^{\mu}
= \frac{\mu}{c}\left(g^{\mu\nu}+\frac{n^2-1}{n^2}\eta^\mu\*\eta^\nu\right)
J_{\nu},\eqno(29)$$
where the corresponding Lorentz gauge is defined in the Schwinger {\it et
al.} (1976) article in the following form

$$\partial_{\mu}A^{\mu} - (\mu\varepsilon-1)(\eta\partial)(\eta A) = 0,
\eqno(30)$$
where $\eta^{\mu} = (1,{\bf 0})$ is the unit time-like vector in the rest
frame of the medium.
The four-potentials are $A^{\mu}(\phi,{\bf A~})$ and
the four-current $J^{\mu}(c\varrho,{\bf J})$,
$n$ is the index of refraction of this medium.

So, the massive electrodynamics in medium can be constructed by generalization
of massless electrodynamics to the case with massive photon.

In superconductiviy, photon is a massive spin 1 particle
as a consequence of a broken symmetry of the
Landau-Ginzburg Lagrangian. The Meissner effect can be used as a
experimental demonstration that photon in a
superconductor is a massive particle.
In particle physics the situation is analogous to the situation in
superconductivity. The masses of particles are also generated by the broken
symmetry or in other words by the Higgs mechanism. Massive particles
with spin 1 form the analogue of the massive photon.

Kirzhnitz and Linde (1972 )proposed a qualitative analysis wherein they
indicated that, as in the Landau-Ginzburg theory of superconductivity,
the Meissner effect can also be realized in the Weinberg model. Later,
it was shown that the Meissner
effect is  realizable in renormalizable gauge fields and also in the Weinberg
model (Yildiz, 1977).

While the photon propagator $D(k)$ in the momentum representation in the
massless electrodynamics is

$$D(k)  =  \frac {1}{|{\bf k}|^2-n^2(k^0)^2-i\epsilon}, \eqno(31) $$
the massive photon propagator is of the form
(here we introduce $\hbar$ and $c$):

$$D(k,m^{2}) =  \frac {1}
{|{\bf k}|^2-n^2(k^0)^2+\frac {m^2\*c^2}{\hbar^{2}}-i\epsilon},
\eqno(32)$$
where this propagator is derived from an assumption that the photon
energetic equation is

$$|{\bf k}|^2 - n^2(k^0)^2 = - \frac {m^2\*c^2}{\hbar^{2}},
\eqno(33)$$
where $n$ is the parameter of the medium and $m$ is mass of
photon in this medium.

From eq. (33) the dispersion law for the massive photons follows:
$$\omega = \frac {c}{n}\sqrt{k^{2} + \frac {m^{2}c^{2}}{\hbar^{2}}}.
\eqno(34)$$

Let us remark here that such dispersion law is valid not only for
the massive photon but also for electromagnetic field in waveguides
and electromagnetic field in ionosphere. It means that the corresponding
photons are also massive and the theory of massive photons is physically
meaningful.

The validity of eq. (33) can be verified using very simple idea
that for $n = 1$ the Einstein equation for mass and energy has to follow.
Putting ${\bf p} = \hbar{\bf k}, \quad \hbar k^{0} =
\hbar (\omega/c) = (E/c)$, we get the Einstein energetic equation

$$E^{2} = {\bf p}^{2}c^{2} + m^{2}c^{4}.
\eqno(35)$$

The propagator for the massive photon is then derived as

$$D_{+}(x-x',m^{2}) = \frac{i}{c}\frac{1}{4\pi^2}
\int_{0}^{\infty}\,d\omega\,\frac{\sin[\frac{n^2\omega^2}{c^2}-
\frac {m^2\*c^2}{\hbar^{2}}]^{1/2}
|{\bf x}-{\bf x}'|}{|{\bf x}-{\bf x}'|}\*e^{-i\omega\*|t-t'|}.
\eqno(36)$$

The function (36) differs from the the original function
$D_{+}$ by the factor

$$\left(\frac {\omega^2\*n^2}{c^2} -
\frac {m^2\*c^2}{\hbar^{2}}\right)^{1/2}.
\eqno(37)$$

From eq. (36) the potentials generated by the massless or massive
photons respectively follow. In case of the massless photon, the potential is
according to Schwinger defined by the formula ($m = 0$):

$$V({\bf x} - {\bf x'}) = \int_{-\infty}^{\infty}d\tau
D_{+}({\bf x - \bf x}',\tau) =
\int_{-\infty}^{\infty}d\tau \left\{
\frac {i}{c}\*\frac {1}{4\pi^2}\*\int_{0}^{\infty}d\omega \frac {\sin
\frac {n\omega}{c}\*|{\bf x}-{\bf x}'|}{|{\bf x}-{\bf x}'|}\*e^{-i\omega|\tau|}
\right\}.
\eqno(38)$$

The $\tau$-integral can be evaluated using the mathematical formula

$$\int_{-\infty}^{\infty}\,d\tau\, e^{-i\omega|\tau|} = \frac {2}{i\omega}
\eqno(39)$$
and the $\omega$-integral can be evaluated using the formula

$$\int_{0}^{\infty}\frac {\sin ax}{x}dx = \frac {\pi}{2}, \quad {\rm for}
\quad a>0.
\eqno(40)$$

After using eqs. (39) and (40), we get

$$V({\bf x} - {\bf x'}) =\frac {1}{c} \frac {1}{4\pi}
\frac {1}{|{\bf x} - {\bf x}'|}.
\eqno(41)$$

In case of the massive photon, the mathematical determination of potential
is the analogical to the massless situation only with the difference we use
the propagator (36) and the table integral (Gradshteyn {\it et al.}, 1965)

$$\int_{0}^{\infty}\frac {dx}{x}\sin\left(p\sqrt{x^{2}-u^{2}}\right) =
\frac {\pi}{2}e^{-pu}.
\eqno(42)$$

Using this integral we get that the potential generated by the massive
photons is

$$V({\bf x} - {\bf x'},m^{2}) = \frac {1}{c}\frac {1}{4\pi}
\frac {\exp{\left\{-\frac {mcn}{\hbar}|{\bf x} - {\bf x'}|\right\}}}
{|{\bf x} - {\bf x}'|}.
\eqno(43)$$

So, we see that in case of the massive photons, the potential generated
by the massive particle is of the Yukawa form.

\section{Volkov solution of the Dirac equation with massless photons}

Let us remember the derivation of the Volkov (1935)
solution of the Dirac
equation in vacuum (we use here the method of derivation and
metric convention of Berestetzkii {\it et al.} (1989)):

$$(\gamma(p-eA) - m)\Psi = 0. \eqno(44)$$

where

$$A^{\mu} = A^{\mu}(\varphi); \quad \varphi = kx.\eqno(45)$$

We suppose that the four-potential satisfies the Lorentz gauge condition

$$\partial_{\mu}A^{\mu} = k_{\mu}\left(A^{\mu}\right)' =
\left(k_{\mu}A^{\mu}\right)' = 0, \eqno(46)$$
where the prime denotes derivation with regard to $\varphi$. From the
last equation follows

$$kA = const = 0,\eqno(47)$$
because we can put the constant to zero. The tensor of electromagnetic field is

$$F_{\mu\nu} = k_{\mu}A'_{\nu} - k_{\nu}A'_{\mu}.\eqno(48)$$

Instead of the linear Dirac equation (44) we consider the quadratical
equation, which we get by multiplication of the linear equation by
operator $(\gamma(p-eA) + m)$, (Berestetzkii {\it et al.}, 1989).
We get:

$$\left[(p - eA)^{2} -m^{2} - \frac{i}{2}eF_{\mu\nu}
\sigma^{\mu\nu}\right]\psi = 0. \eqno(49)$$

Using $\partial_{\mu}(A^{\mu}\psi) = A^{\mu}\partial_{\mu}\psi$, which
follows from eq. (46), and $\partial_{\mu}\partial^{\mu} =
\partial^{2} = -p^{2}
$, with $p_{\mu} = i(\partial /\partial x^{\mu}) = i\partial_{\mu}$, we get
the quadratical Dirac equation for the four potential of the plane wave:

$$[-\partial^{2} - 2i(A\partial) + e^{2}A^{2} - m^{2} -
ie(\gamma k)(\gamma A')]\psi = 0. \eqno(50)$$

We are looking the solution of the last equation in the form:

$$\psi = e^{-ipx}F(\varphi).\eqno(51)$$

After insertion of this equation into (50), we get with ($k^{2} = 0$)

$$\partial^{\mu}F = k^{\mu}F', \quad \partial_{\mu}\partial^{\mu}F = k^{2}F''
= 0,\eqno(52)$$
the following equation for $F(\varphi)$

$$2i(kp)F' + [-2e(pA) + e^{2}A^{2} - ie(\gamma k)(\gamma A')]F = 0. \eqno(53)$$

The integral of the last equation is of the form:

$$F = \exp\left\{-i\int_{0}^{kx}\left[\frac {e(pA)}{(kp)} - \frac
{e^{2}}{2(kp)}A^{2}\right]
d\varphi + \frac {e(\gamma k)(\gamma A)}{2(kp)}\right\}
\frac{u}{\sqrt{2p_{0}}}, \eqno(54)$$
where $u/\sqrt{2p_{0}}$ is the arbitrary constant bispinor.

Al powers of $(\gamma k)(\gamma A)$ above the first are equal to zero,
since

$$(\gamma k)(\gamma A)(\gamma k)(\gamma A) =
- (\gamma k)(\gamma k)(\gamma A)(\gamma A) +
2(kA)(\gamma k)(\gamma A) = -k^{2}A^{2} = 0.\eqno(55)$$

Then we can write:

$$\exp\left\{e\frac {(\gamma k)(\gamma A)}{2(kp)}\right\} =
1 + \frac {e(\gamma k)(\gamma A)}{2(kp)}.\eqno(56)$$

So, the  solution is of the form:

$$\Psi_{p} = R \frac {u}{\sqrt{2p_{0}}}e^{iS}  =
\left[1 + \frac {e}{2kp}(\gamma k)(\gamma A)\right]
\frac {u}{\sqrt{2p_{0}}}e^{iS},
\eqno(57)$$
where $u$ is an electron bispinor of the corresponding Dirac equation

$$(\gamma p - m)u = 0.\eqno(58)$$

The mathematical object $S$ is the classical Hamilton-Jacobi function,
which  was determined in the form:

$$S = -px - \int_{0}^{kx}\frac {e}{kp}\left[(pA) - \frac {e}{2}
(A)^{2}\right]d\varphi. \eqno(59)$$

The current density is

$$j^{\mu} = {\bar \Psi}_{p}\gamma^{\mu}\Psi_{p},
\eqno(60)$$
where $\bar\Psi$ is defined as the transposition of (57), or,

$$\bar\Psi_{p} = \frac {\bar u}{\sqrt{2p_{0}}}\left[1 +
\frac {e}{2kp}(\gamma A)(\gamma k)\right]
e^{-iS}.
\eqno(61)$$

After insertion of $\Psi_{p}$ and $\bar\Psi_{p}$
into the current density, we have:

$$j^{\mu} = \frac {1}{p_{0}}\left\{p^{\mu} - eA^{\mu} +
k^{\mu}\left(\frac {e(pA)}{(kp)} - \frac {e^{2}A^{2}}{2(kp)}\right)
\right\}.
\eqno(62)$$
which is in agreement with formula in the Meyer article (1971).

The so called kinetic momentum corresponding to $j^{\mu}$ is as follows:

$$J^{\mu} = \Psi^{*}_{p}(p^{\mu} - eA^{\mu})\Psi_{p})
= {\bar \Psi}_{p}\gamma^{0}(p^{\mu} - eA^{\mu})\Psi_{p}) = $$

$$\left\{p^{\mu} - eA^{\mu} +
k^{\mu}\left(\frac {e(pA)}{(kp)} - \frac {e^{2}A^{2}}{2(kp)}\right)
\right\} + k^{\mu}\frac
{ie}{8(kp)p_{0}}F_{\alpha\beta}(u^{*}\sigma^{\alpha\beta}u),
\eqno(63)$$
where

$$\sigma^{\alpha\beta} = \frac {1}{2}(\gamma^{\alpha}\gamma^{\beta} -
\gamma^{\beta}\gamma^{\alpha}).\eqno(64)$$

\section{Volkov solution of the Dirac equation for massive photons}

The original Volkov solution
is based on the assumption that photon has zero rest mass, or, $k^{2}= 0$.
Our goal is to consider the solution of the Dirac equation in case that
$k^{2} = M^{2}$, where $M$ is the rest mass of photon. We use here the
metrical notation of Berestetzkii {\it et al.} (1989).

We apply the procedure of the preceding
section for  the case of the massive photon, and
we write:

$$\psi = e^{-ipx}F(\varphi),\eqno(65)$$
where for $F$ we get the following equation

$$M^{2}F'' - 2i(kp)F' + G(\varphi)F = 0\eqno(66)$$
with

$$G(\varphi) = 2e(pA) - e^{2}A^{2} + ie(\gamma k)(\gamma A').\eqno(67)$$

The equation (66) differs from the original
Volkov equation only by means of the massive
term. However, the equation is substantially new, because of the second
derivative of the function $F$. The solution of the last equation can be
easily obtained in the approximative form in case that $M \rightarrow 0$.
However, let us try to find  the exact solution, which was not
described , to our in physical or mathematical journals.

In order to find such  solution, we transcribe this equation in the
form:

$$F'' + aF' + bF = 0,\eqno(68)$$
where

$$a = - \frac {2i(kp)}{M^{2}}, \quad b(\varphi) = \frac {G(\varphi)}{M^{2}}.
\eqno(69$$

Using the substitution

$$F = v(\varphi)e^{-\frac {1}{2}a\varphi},\eqno(70)$$
we get simple equation for $v(\varphi)$:

$$v'' + P(\varphi)v = 0,\eqno(71)$$
where

$$P(\varphi) = (-\frac {a^{2}}{4} + b).\eqno(72)$$

Using the substitution

$$v(\varphi) = e^{\int_{0}^{\varphi}T(\varphi)d\varphi},\eqno(73)$$
we get from eq. (71)
$$T' + T^{2} + P(\varphi) = 0.\eqno(74)$$

Equation (74) is so called Riccati equation. The mass term is hidden 
in $P(\varphi)$.
It is well know
that there is no general form of solution of this equation. There is only
some solution expressed in the elementary functions for some specific
functions $P(\varphi)$. Nevertheless, there is interesting
 literature concerning
the Riccati equation. For instance, Riccati equation is applied in the
supersymmetric quantum mechanics (Cooper {\it et al.}, 1995), in variational
calculus (Zelekin, 1998), nonlinear physics, (Matveev {\it et al.}, 1991), in
renormalization group theory  (Buchbinder {\it et al.}, 1992; Milton {\it et al.},
2001) and in thermodynamics (Rosu {\it et al.}, 2001).

With regard to circumstances, we are forced to find
some approximative solution with form similar to the original Volkov solution.
Let us show the derivation of such the approximative solution. We hope it will
play the same role in quantum electrodynamics with the massive photon
as in the case with the massless photon.

There are many approximative methods for solution of this problem. We
choose the elementary method which was also applied to the
 Schr$\ddot{\rm o}$dinger
equation and which is  described for instance in the monograph
of Mathews {\it et al.} (1964).

The approximation consists at the application of the following inequalities:

$$|F''(\varphi)| \ll |F'(\varphi)|; \quad  |F''(\varphi)| \ll
 |F(\varphi)|.
 \eqno(75)$$

Then, we get the original Volkov solution with the difference that
the existence of the nonzero photon mass will be involved only in the
exponential expansion. Or, with $U = e(\gamma k)(\gamma A)/2(kp)$, we perform
the expansion:

$$e^{U} = \left\{ 1 + \frac {1}{1!}U + \frac {1}{2!}U^{2} +
\frac {1}{3!}U^{3} + ... \right\} = $$

$$\left\{1 + \frac {e}{2(kp)}(\gamma k)(\gamma A) + \frac {1}{2!}
\left(\frac {e}{2(kp)}\right)^{2}(-M^{2}A^{2}) + \right.$$

$$\left.\frac {1}{3!}
\left(\frac {e}{2(kp)}\right)^{3}(\gamma k)(\gamma A)(-M^{2}A^{2}) +
 \frac {1}{4!}\left(\frac {e}{2(kp)}\right)^{4}(M^{4}A^{4}) + ...
\right\},\eqno(76)$$
where we have used equation (55) in the modified form

$$(\gamma k)(\gamma A)(\gamma k)(\gamma A) =
- (\gamma k)(\gamma k)(\gamma A)(\gamma A) +
2(kA)(\gamma k)(\gamma A) = -k^{2}A^{2} = -M^{2}A^{2}.\eqno(77)$$
with  $k^{2} = M^{2}$ for massive photons. We see that in this method of
approximation the Massive solution involves the Volkov solution as the
basic term and then the additional terms containing photon  mass.

After performing some algebraic operations, we get the first approximation of
the Volkov solution with the massive photon in the following form

$$\Psi_{p} = R(A, M^{2}) \frac {u}{\sqrt{2p_{0}}}e^{iS}  =
\left[1 + \frac {e}{2kp}(\gamma k)(\gamma A) -
\left(\frac {e}{2kp}\right)^{2}M^{2}A^{2} + ... \right]
\frac {u}{\sqrt{2p_{0}}}e^{iS}.\eqno(78)$$

Now, we are prepared to solve some physical problems with the Volkov
solution with massive photons.

\section{Emission of massive photons by electron moving in the  periodic
field}

Let us consider the monochromatic circularly
polarized electromagnetic wave with the four potential

$$A = a_{1}\cos\varphi + a_{2}\sin\varphi;\quad a_{3} = 0; \quad \varphi = kx \eqno(79)$$
with $k^{\mu} = (\omega, {\bf k})$ being a wave 4-vector and $k^{2} = M^{2}$,
the 4-amplitudes $a_{1}$ and $a_{2}$ are the same and one another
perpendicular, or

$$a_{1}^{2} = a_{2}^{2} = a^{2}; \quad  a_{1}a_{2} = 0.\eqno(80)$$

We shall also use the Lorentz gauge condition, which gives
$a_{1}k = a_{2}k = 0$.

The wave function is then of the form:

$$\psi_{p} = \left\{1 + \frac {e}{2(kp)}\left[(\gamma k)(\gamma a_{1})
\cos\varphi + (\gamma k)(\gamma a_{2})\sin\varphi -
\frac {e}{2(kp)}a^{2}M^{2} + ...\right]\right\} \frac {u(p)}{\sqrt{2q_{0}}}
\times $$

$$\exp\left\{-ie\frac {a_{1}p}{(kp)}\sin\varphi +
ie\frac {a_{2}p}{(kp)}\cos\varphi  -iqx\right\},\eqno(81)$$
where

$$q^{\mu} = p^{\mu} - e^{2}\frac {a^{2}}{2(kp)}(k^{\mu})\eqno(82)$$
is the time-averaged value of the  eq. (62).

The corresponding matrix element is of he obligate form
(Berestetzkii {\it et al.}, 1989).

After performing the appropriate mathematical operation we get the
$\delta$-function in the matrix element, from which the conservations
laws follow in the form

$$sk + q = q' + k'\eqno(83)$$.

The interpretation of this formula is as follows: $s$ massive photons
with momentum $k$ are absorbed by electron with momentum $q$
and only one massive photon is emitted  with the 4-vector $k'$,
and the final momentum of electron is $q'$. So, we see that the Volkov
solution gives the multiphoton processes, which are intensively
studied in the modern physics (Delone {\it et al.}, 2000).

For the periodic wave it is

$$q^{2} = q'^{2} = m_{*}^{2}; \quad
m_{*} = m\; \sqrt{1  + \frac {e^{4}a^{4}}{(2kp)^{2}}\frac {M^{2}}{m^{2}} -
\frac {e^{2}}{m^{2}} a^{2}}\eqno(84)$$
which can be interpreted as a mass shift of electron in the periodic
field, or, the mass renormalization.

If we consider an electron at a rest $({\bf q} = 0, q_{0} = m_{*})$,
 then from the formula (82), (83) and (84) follows

$$(s^2 + 1)\frac{M^2}{2m_{*}}\frac{1}{\omega\omega'} + s\frac {1}{\omega'}
- \frac {1}{\omega} = \frac {s}{m_{*}}(1 - \cos\Theta);
\quad s = 1, 2, 3, ...\; n. \eqno(85)$$

The massless limit of the last formula is the well known Compton
formula (with M = 0)

$$\omega' = \frac {s\omega}{1 + \frac {s\omega}{m_{*}}(1 - \cos\theta)},
\eqno(86)$$
where $\theta$ is an angle between ${\bf k}$ and ${\bf k}'$. So we see
that frequencies $\omega'$ are harmonic frequencies of $\omega$.

\section{Emission of massive photons by electron moving
in the impulsive force}

We use the $\delta$-function form of the ultrashort laser pulse (Pardy, 2003)

$$A_{\mu} = a_{\mu}\eta(\varphi), \eqno(87)$$
where $\eta(\varphi)$ is the Heaviside unit step function defined as
follows: $\eta(\varphi) = 0; \varphi < 0 $, and $\eta(\varphi) = 1;
\varphi \geq 0$.  Then, the function $S$ and $R$ in the Volkov
solution
$\Psi_{p}$ are as follows (Pardy, 2003):

$$S = -px - \left[e\frac {ap}{kp} - \frac {e^{2}}{2kp}a^{2}\right]\varphi,
\quad R = \left[1 + \frac {e}{2kp}(\gamma k)(\gamma a)\eta(\varphi) +
...\right].\eqno(88)$$

So, we get the matrix element in the form:

$$M = g \int d^{4}x\bar\Psi_{p'}O\Psi_{p}\frac {e^{ik'x}}{\sqrt{2\omega'}},
\eqno(89)$$
where  $O = \gamma e'^{*}$, $g = -ie^{2}$ in case of the
electromagnetic interaction and

$$\bar \Psi_{p'} = \frac {\bar u}{\sqrt{2p'_{0}}}\bar R(p')e^{-iS(p')}.
\eqno(90)$$

In such a way, using above definitions, we write the matrix element in the
form:

$$M = \frac{g}{\sqrt{2\omega'}} \frac {1}{\sqrt{2p'_{0}2p_{0}}}
\int d^{4}x\bar R(p')OR(p)e^{-iS(p') + iS(p)} e^{ik'x}.
\eqno(91)$$

The quantity $\bar R(p')$ follows immediately from eq. (87), namely:

$$\bar R' =\overline
{\left[1 + \frac {e}{2kp'}(\gamma k)(\gamma a)\eta(\varphi) + ...\right]}  =
{\left[1 + \frac {e}{2kp'}(\gamma a)(\gamma k)\eta(\varphi) + ...\right]}.
\eqno(92)$$

Using

$$-iS(p') + iS(p) = i(p'-p) + i(\alpha' - \alpha)\varphi, \eqno(93)$$
where

$$\alpha = \left(e\frac {ap}{kp} - \frac {e^{2}}{2}\frac {a^{2}}{kp}\right),
\quad \alpha' = \left(e\frac {ap'}{kp'} - \frac {e^{2}}{2}
\frac {a^{2}}{kp'}\right),\eqno(94)$$
we get:

$$M = \frac{g}{\sqrt{2\omega'}} \frac {1}{\sqrt{2p'_{0}2p_{0}}}
\int d^{4}x\bar u(p')\bar R(p')OR(p)u(p)e^{i(p' - p)x}
e^{i(\alpha' - \alpha)\varphi}e^{ik'x}.
\eqno(95)$$

We get after $x$-integration:

$$M = \frac{g}{\sqrt{2\omega'}} \frac {1}{\sqrt{2p'_{0}2p_{0}}}
\bar u(p')R(p')OR(p)u(p)\delta^{(4)}(kl + p - k' - p').
\eqno(96)$$

We see from the presence of the $\delta$-function in eq. (96) that during
the process of the interaction of electron with the laser pulse the
energy-momentum conservation law holds good:

$$lk + p = k' + p'; \quad l = \alpha-\alpha'. \eqno(97)$$

The last equation describes the so called multiphoton process, which
can be also described using Feynman diagrams and which are studied in
the different form intensively in the modern physics of multiphoton
ionization of atoms (Delone {\it et al.}, 2000; Pardy, 2003).

If we introduce the angle $\Theta$ between ${\bf k}$ and ${\bf k}'$, then,
with  $|{\bf k}| = \omega$  and  $|{\bf k}'| = \omega'$, we get from
the squared equation (97) in the rest system of electron, where
$p = (m_{*},0)$, the following equation $k = (\omega, {\bf k})$:

$$(l^2 + 1)\frac{M^2}{2m_{*}}\frac{1}{\omega\omega'} + l\frac {1}{\omega'}
- \frac {1}{\omega} = \frac {l}{m_{*}}(1 - \cos\Theta);
\quad l = \alpha - \alpha', \eqno(98)$$
which is modification of the original equation for the Compton process

$$\frac {1}{\omega'} - \frac {1}{\omega} = \frac {1}{m}(1 - \cos\Theta).
\eqno(99)$$

We see that the substantial difference between single photon interaction
and $\delta$-pulse interaction is the factor $l = \alpha - \alpha'$.

We know that the last formula of the original Compton effect can be written
in the form suitable for the experimental verification, namely:

$$\Delta \lambda = 4\pi\frac{\hbar}{mc}\sin^{2}\frac {\Theta}{2},
\eqno(100)$$
which was used by Compton for the verification of the quantum
nature of light (Rohlf, 1994).

Let us remark, the equation $lk + p = k' + p'$ is the symbolic
expression of the nonlinear Compton effect and it concerns
only the situation where $l$ photons are absorbed at a single point,
and it does not describe the  process  where electron scatters twice,
or more, as it traverses the laser focus.
The nonlinear Compton process was experimentally confirmed (Bulla {\it et
al.}, 1996).

\section{Discussion}
The present article is continuation of the author
discussion on laser interaction with electrons (Pardy, 1998, 2001),
where the Compton model of laser acceleration was proposed and author article (Pardy, 2003),
where the $\delta$-form laser pulse was considered.

The $\delta$-form laser pulses are the idealization
of the experimental situation in laser physics. It was
demonstrated theoretically that at present time the zeptosecond and
subzeptosecond laser pulses of duration $10^{-21} - 10^{-22}$ s can be
realized by the petawat lasers (Kaplan {\it et al.}, 2002).
It means that the generation of
the ultrashort laser pulses is the keen interest
in development of laser physics.

We have derived  modified Compton formulas which involve multiphoton
interaction of laser beam  with electron. In case of the periodic field,
the multiplicity is formed by the natural numbers and in case of the
$\delta$-pulse, by number $l = \alpha - \alpha'$. This effect
can be interpreted in such a way that the photonic object with $s$ or $l$
photons interacts simultaneously with one electron. We do not think that
the photonic object is consequence of the Bose-Einstein
condensation of photons in laser beam.
It behaves as photonic elementary object
and probably it can be used in the experiments in particle physics.

The Volkov solution of the Dirac equation  for electromagnetic
potential with massive photons concerns
not only the  superconductive medium but also the
electron-positron
plasma, ionosphere medium, photons in waveguides, or massive photons generated
hypothetically during inflation (Prokopec {\it et al.}, 2003).

The bosons
$W^{\pm}$ and $Z^{0}$ are also massive and it means that the generalization
of our approach to the situation in the standard model is evidently
feasible. The vector mesons
$\rho, \varphi,J/\psi$ are generated during the nuclear collisions
and probably, the Volkov solution for these massive vector particles
will play substantial role in the nuclear physics.

\vspace{11mm}

{\bf References}

\bigskip

\noindent
Anderson, P. W.(1963). Plasmons, gauge invariance, and mass,
{\it Physical Review} {\bf 130}(1), 439.\\[2mm]
Berestetzkii, V. B., Lifshitz, E. M. and Pitaevskii, L. P. (1989). \\
{\it Quantum Electrodynamics}, Moscow, Nauka. (in Russian). \\[2mm]
Bulla, C. {\it et al.} (1996). Observation of nonlinear effects in Compton
scattering, {\it Physical Review Letter} {\bf 76}, 3116. \\[2mm]
Buchbinder, I. L., Odintsov, S. D. and  Shapiro, I. L. (1992).
 {\it Effective Action in Quantum Gravity}, IOP Publishing Ltd.\\[2mm]
Commins, E. D. and Bucksbaum, P. H. (1983). {\it Weak Interactions of Leptons
and Quarks}, Cambridge University Press, Cambridge.\\[2mm]
Cooper, F.,  Khare, A. and Sukhatme, U. (1995).  Supersymmetry and
 quantum mechanics,
{\it Physics Reports} {\bf 251}, 267. \\[2mm]
Delone, N. B. and Krainov, V. P. (2000). {\it Multiphoton Processes in Atoms},
2nd ed., Springer-Verlag, Berlin, Heidelberg, New York.\\[2mm]
de Wit B.  and Smith, J. (1986). {\it Field Theory in Particle Physics},
Vol. I, Elsevier. \\[2mm]
Dittrich, W. (1978).  Source methods in quantum field theory,
{\it Fortschritte der Physik} {\bf 26},  289.\\[2mm]
Feldman, G. and Mathews, P. T. (1963). Massive electrodynamics,
{\it Physical Review} {\bf 130}, 1633.\\[2mm]
Goldhaber, A. S. and Nieto, M. M. (1971).
Terrestrial and extraterrestrial limits on the photon mass,
{\it Review of Modern Physics} {\bf 43}, 277.\\[2mm]
Gradshteyn, I. S. and Ryzhik, I. M.  (1965). {\it Table of Integrals,
Series and Products} Academic Press, New York. \\[2mm]
Kaplan, A. E. and Shkolnikov, P. L. (2002). Lasetron: A proposed source of
powerful nuclear-time-scale electromagnetic bursts,
{\it Physical Review Letter} {\bf 88}(7),  074801.\\[2mm]
Kirzhnitz, A. D. and Linde, A. D. (1972). Macroscopic consequences
of the Weinberg model,
{\it Physics Letters} {\bf 42B}, 471. \\[2mm]
Matveev, V. B. and  Salle, M. A. (1991). {\it Darboux Transformations
and Solitons} Springer, Berlin.\\[2mm]
Mathews, J. and Walker, R. L. (1964). {\it Mathematical Methods of Physics}
W. B. Benjamin, Inc., New York -- Amsterdam. \\[2mm]
Meyer, J. W. (1971). Covariant classical motion of electron in a laser beam,\\
{\it Physical Review D: Particles and Fields} {\bf 3}(2), 621. \\[2mm]
Milton, K., Odintsov, S. D.  and Zerbini, S. (2001).  Bulk versus
brane running couplings, e-print hep-th/0110051 \\[2mm]
Minkowski, P. and Seiler, R. (1971). Massive vector meson in external fields,
{\it Physical Review  D: Particles and Fields} {\bf 4}, 359. \\[2mm]
Okun, L. B. (1981). {\it Leptons and Quarks}, Nauka, Moscow.
(in Russian).\\[2mm]
Okun, L. B. (2002). Photons, clocks, gravity and the concept of mass,
{\it Nuclear  Physics B} (Proc. Supl.) {\bf 110}, 151. \\[2mm]
Okun, L. B., Selivanov, K.  and Telegdi, V. (1999). Gravitation,
photons, clocks, {\it Physics Uspekhi} {\bf 42}, 1045.\\[2mm]
Pardy, M. (1994). The \v Cerenkov effect with radiative corrections,
{\it Physics  Letters B} {\bf 325}, 517.\\[2mm]
Pardy, M. (1998). The quantum field theory of laser acceleration,
{\it Physics  Letters A} {\bf 243},  223. \\[2mm]
Pardy, M. (2001). The quantum electrodynamics of laser acceleration,
{\it Radiation Physics and Chemistry} {\bf 61},  391.\\[2mm]
Pardy, M. (2002). \v Cerenkov effect with massive photons,
{\it International Journal of Theoretical Physics} {\bf 41}(5), 887.\\[2mm]
Pardy, M. (2003). Electron in the ultrashort laser pulse,
{\it International Journal of Theoretical Physics} {\bf 42}(1), 99.\\[2mm]
Prokopec, T. and Woodard, D. (2003). Vacuum polarization and photon
mass in inflation, e-print astro-ph/0303358; CERN-TH/2003-065.\\[2mm]
Ritus, V. I. (1969). Radiative effects and their enhancement in
intensive electromagnetic field, {\it Journal of Experimental and Theoretical
Physics} {\bf 57}(6), 2176. (in Russian). \\[2mm]
Rohlf, J. W. (1994). {\it Modern Physics from $\alpha$ to $Z^{0}$}, John
Wiley \& Sons, Inc. New York.
Rosu, A. C. and Aceves de la Crus. (2001). One-parameter Darboux-transformed
quantum action in thermodynamics, e-print quant-ph/0107043 \\[2mm]
Ryder, L. H. (1985).{\it Quantum Field Theory}, Cambridge University Press,
Cambridge. \\[2mm]
Schwinger, J. (1970). {\it Particles, Sources and Fields}, Vol. I
Addison-Wesley, Reading, Massachusets.\\[2mm]
Schwinger, J., Tsai W. Y.  and Erber T. (1976). Classical and quantum theory
of synergic synchrotron-\v Cerenkov radiation,
{\it Annals of Physics (New York)} {\bf 96} , 303. \\[2mm]
Sommerfeld, A. (1954). {\it Optics}, Academic Press, New york 10,
N. Y. USA.\\[2mm]
van Nieuwenhuizen, P. (1973). Radiation of massive gravitation,
{\it Physical Review D: Particles and Fields} {\bf 7}(8), 2300. \\[2mm]
Volkov, D. M., (1935).  $\ddot{\rm U}$ber eine Klasse
von L$\ddot{\rm o}$sungen
der Diracschen Gleichung, {\it Zeitschrift f$\ddot{\it u}$r Physik},
{\bf 94}. 250\\[2mm]
Yildiz, A. (1977). Meissner effect in gauge fields,
{\it Physical Review D: Particles and Fields} {\bf 16}(12), 3450.\\[2mm]
Zelekin, M. I. (1998). {\it Homogenous Spaces and Riccati Equation in
Variational Calculus}, Factorial, Moscow. (in Russian).

\end{document}